\documentclass[superscriptaddress,aps,preprintnumbers,amsmath,showpacs,amssymb,prd,nofootinbib,preprint,longbibliography]{revtex4-1}
\pdfoutput=1
\usepackage{booktabs} 
\usepackage[normalem]{ulem}
\usepackage{array} 
\usepackage[table,xcdraw]{xcolor} 
\usepackage{amsmath,amssymb}
\usepackage{graphicx}
\usepackage{fancyhdr}
\usepackage{bm, color}
\usepackage{slashed,amsthm,amsfonts,empheq}
\usepackage[caption=false]{subfig}
\usepackage{hyperref}
\usepackage{booktabs}
\usepackage{multirow}
\usepackage[left=2cm,right=2cm,top=2cm,bottom=2cm,includefoot,a4paper]{geometry}
\usepackage{tikz}
\usetikzlibrary{arrows.meta}

\newcommand{\Slash}[1]{{\ooalign{\hfil#1\hfil\crcr\raise.167ex\hbox{/}}}}

\newcommand{\beq}{\begin{equation}}  \newcommand{\eeq}{\end{equation}}
\newcommand{\bef}{\begin{figure}}  \newcommand{\eef}{\end{figure}}
\newcommand{\bec}{\begin{center}}  \newcommand{\eec}{\end{center}}

\newcommand{\laq}[1]{\label{eq:#1}}  
\newcommand{\Eq}[1]{Eq.(\ref{eq:#1})}
\newcommand{\Eqs}[1]{Eqs.(\ref{eq:#1})}
\newcommand{\eq}[1]{(\ref{eq:#1})}

\newcommand{\lac}[1]{\label{chap:#1}}

\def\({\left(}
\def\){\right)}

\newcommand{\GEV}{\,{\rm GeV}}

\def\F{\Phi}

\def\*{\dagger}

\begin{document}

\title{\Large Earth-scale searches for displaced vertices: \\ KM3NeT meets the LHC}

\author{Ui Min}
\affiliation{Department of Physics, Tokyo Metropolitan University, Minami-Osawa, Hachioji-shi, Tokyo 192-0397, Japan}

\author{Wen Yin}
\affiliation{Department of Physics, Tokyo Metropolitan University, Minami-Osawa, Hachioji-shi, Tokyo 192-0397, Japan}

\begin{abstract}
At high-energy colliders, enormous numbers of feebly interacting particles beyond the Standard Model can be produced with negligible missing-energy signatures in the main detectors. These particles can travel macroscopic distances and decay inside a {distant} large-volume detector, leaving visible signals. In this sense, we may not need new detectors to search for very long-lived particles.
In this paper, we show that long-lived dark particles produced in rare meson decays in proton--proton collisions during {LHC Run~3 and at the HL-LHC} 
can leave observable imprints in KM3NeT detectors such as 
ORCA. {If no anomalous signal has been observed in ORCA to date, our projected sensitivity for LHC Run~3 will correspond to one of the strongest limits on kaon decays into light long-lived particles.}
\end{abstract}

\maketitle

\section{Introduction}

After the discovery of the Higgs boson at the LHC, particles beyond the Standard Model (BSM) are still being searched for. 
Given the upcoming HL-LHC and possible future colliders such as FCC-ee, FCC-hh, and ILC, it may be timely to consider how to fully exploit the enormous number of collisions provided by these machines for generic BSM searches \cite{Apollinari:2017cqg,Abada:2019lih}. 

\begin{figure*}[htbp]
\centering
  \includegraphics[
    width=0.9\linewidth,
    keepaspectratio
  ]{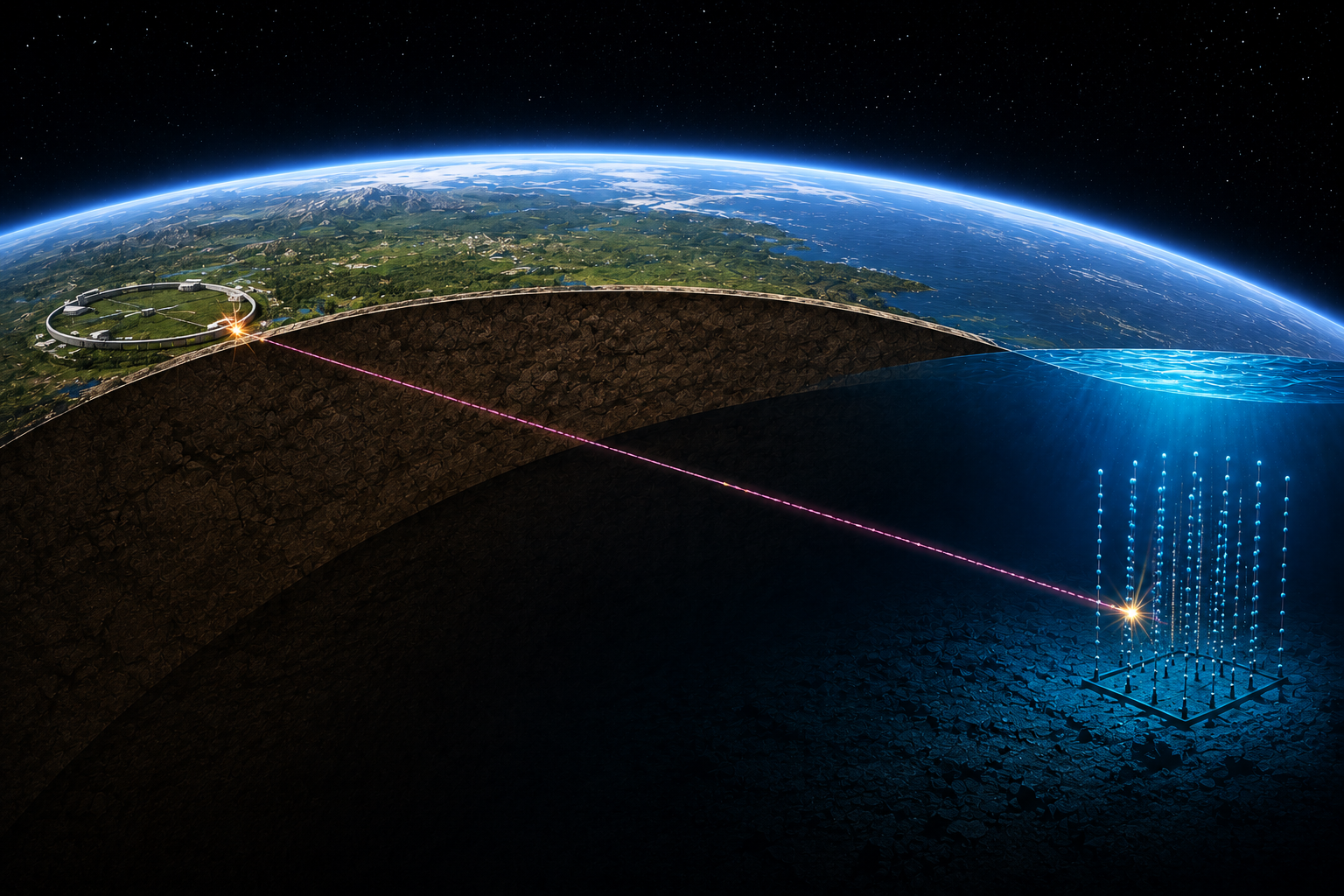}
\caption{Schematic of an Earth-scale displaced vertex in KM3NeT-ORCA for a long-lived dark particle produced at the LHC. 
}
\label{fig:image_geometric_detection}
\end{figure*}

In this paper, we point out a new strategy for searching for BSM particles, especially light and long-lived hidden particles that can be produced efficiently \cite{Nakayama:2014cza,Takahashi:2020uio, Sakurai:2021ipp,Haghighat:2022qyh,Yin:2024txg,Albertus:2026fbe}, by using a large-volume detector located far away from the collider (see Fig.~\ref{fig:image_geometric_detection}): namely, a search for geometrically long displaced vertices.

Various dedicated detectors have been proposed to search for long-lived particles produced at colliders, including FASER, MATHUSLA, and CODEX-b \cite{Feng:2017uoz,Chou:2016lxi,Gligorov:2017nwh}; searches using large-volume neutrino detectors for long-lived particles produced in atmospheric showers have also been considered \cite{Arguelles:2019ziu,Kachelriess:2021man}. 
Here, instead, we consider whether an existing large-volume detector located hundreds of kilometers away can accidentally serve as a far detector for particles produced at a high-energy collider and therefore provide {complementary information for searches for long-lived BSM particles}. A related idea for very light particles at synchrotron radiation facilities without occupying the beamline was studied in \cite{Yin:2024rjb,Yin:2025awb,Yin:2025bui}. 
We study the relevant geometry, possible production processes of the BSM particles, and the resulting sensitivity. 

The enormous number of mesons produced in hadronic collisions can {in turn} produce light long-lived BSM particles through rare decays.
If their interactions with the Standard Model are sufficiently weak, such particles effectively behave as missing energy in conventional collider experiments and can escape the main detectors before decaying. 
We show that, while satisfying existing limits on invisible meson decays, such particles can survive over macroscopic distances and decay visibly in a {distant} large-volume detector such as KM3NeT/ORCA, which is designed for high-energy neutrino detection \cite{Adrian-Martinez:2016fdl}.
In particular, rare kaon decays into invisible final states, or into a pion plus a missing particle, can provide new limits on BSM particles whose lab-frame decay length is around $400\,{\rm km}$ using existing LHC data.

\section{Basics and Geometry}
\lac{geometry}

Before discussing production and detection in detail, we first present the basic formulas independently of the production mechanism.
Let $X$ be a neutral long-lived particle produced at a collider or fixed-target experiment with energy $E_X$, mass $m_X$, and rest-frame lifetime $\tau_X$.\footnote{{Alternatively, we may consider the scattering of $X$ off particles in the detector. This, however, may not be very promising because $X$ may then be screened by the Earth or other environments through the same interaction. In the main discussion, especially for the meson-decay case, the screening effect is negligible in various models.}}
Its lab-frame decay length is
\begin{equation}
  \lambda_X(E_X)=\beta_X\gamma_X c\tau_X .
  \laq{lambdaX}
\end{equation}
Here $\gamma_X=E_X/m_X$ and $\beta_X=\sqrt{1-\gamma_X^{-2}}$ are the Lorentz factor and velocity, respectively.

For a detector with projected area $A_\perp$ along the line of sight and path length $\ell_{\rm det}$ inside the instrumented volume, the probability for an emitted $X$ to enter the detector and decay visibly inside it is
\begin{equation}
  P_X
  =
  \frac{A_\perp}{L^2}\partial_{\Omega} \F_X
  \exp\left[-\frac{L}{\lambda_X}\right]
  \left\{1-\exp\left[-\frac{\ell_{\rm det}}{\lambda_X}\right]\right\} \simeq \partial_{\Omega} \F_X \frac{V_{\rm eff}}{ L^2\lambda_X}   \exp\left[-\frac{L}{\lambda_X}\right],
    \laq{probvolume}
\end{equation}
where $\partial_\Omega \F_X$ is the angular differential flux of $X$ produced {per collision}. The {approximation} is valid when $\ell_{\rm det}\ll\lambda_X$ and $\ell_{\rm det}\ll L$.
Eq.~\eq{probvolume} is maximized as a function of $\lambda_X$ at
$
  \lambda_X\simeq L,
  $ 
with
\begin{equation}
  P_X^{\rm max}
  \simeq  
  \frac{e^{-1} V_{\rm eff}}{4\pi L^3}\times \left(\frac{\partial_{\Omega} \F_X}{(4\pi)^{-1}}\right).
  \laq{pmax}
\end{equation}
Here $V_{\rm eff}=A_\perp\ell_{\rm det}$ is the effective detector volume, so the optimized probability is proportional to the detector volume.
Thus, even at a very large distance, a sufficiently large detector can be sensitive to a long-lived particle with an appropriate decay width. We write the required number of produced $X$ particles for a target yield $N_X^{\rm ev}$ as $N_X^{\rm req}\equiv N_X^{\rm ev}/P_X^{\rm max}$.

Table~\ref{tab:far_detector_comparison} compares several geometries and gives the required number of produced dark particles, $N_X^{\rm req}$, for three detector events using $P_X^{\rm max}$. Here we assume $\partial_\Omega\F_X=(4\pi)^{-1}$.
The background analysis and production yields are discussed below.
Because the CERN--KM3NeT configurations give the smallest values of $N_X^{\rm req}$, we focus on this geometry. Its details are summarized in Table~\ref{tab:geometry}.

\begin{table}[t]
\centering
\begin{tabular}{lcccc}
\hline\hline
Source $\to$ detector
& Baseline $L$
& $V_{\rm eff}$ benchmark
& $P^{\rm max}_X$
& $N_X^{\rm req}$ for 3 events
\\
\hline
CERN $\to$ KM3NeT/ARCA
& $1.39\times10^3\,{\rm km}$
& $1\,{\rm km}^3$
& $1.1\times10^{-11}$
& $2.7\times10^{11}$
\\
CERN $\to$ KM3NeT/ORCA
& $3.82\times10^2\,{\rm km}$
& $7\times10^{-3}\,{\rm km}^3$
& $3.7\times10^{-12}$
& $8.1\times10^{11}$
\\
SuperKEKB $\to$ Hyper-K
& $\sim2.5\times10^2\,{\rm km}$
& $1.9\times10^{-4}\,{\rm km}^3$
& $3.5\times10^{-13}$
& $8.5\times10^{12}$
\\
SuperKEKB $\to$ Super-K
& $\sim2.5\times10^2\,{\rm km}$
& $2.25\times10^{-5}\,{\rm km}^3$
& $4.2\times10^{-14}$
& $7.1\times10^{13}$
\\
CERN $\to$ IceCube
& $\sim1.18\times10^4\,{\rm km}$
& $1\,{\rm km}^3$
& $1.8\times10^{-14}$
& $1.7\times10^{14}$
\\
Fermilab $\to$ IceCube
& $\sim1.16\times10^4\,{\rm km}$
& $1\,{\rm km}^3$
& $1.9\times10^{-14}$
& $1.6\times10^{14}$
\\
Fermilab $\to$ DUNE FD
& $\sim1.3\times10^3\,{\rm km}$
& $4\times10^{-5}\,{\rm km}^3$
& $5.5\times10^{-16}$
& $5.4\times10^{15}$
\\
\hline\hline
\end{tabular}
\caption{
Comparison of accidental far-detector configurations. The optimized probability is estimated as
$P_{\rm max}=e^{-1}V_{\rm eff}/(4\pi L^3)$, assuming isotropic emission and a lab-frame decay length optimized at $\lambda_X=L$. The required number of produced long-lived particles is $N_X^{\rm req}=3/P_{\rm max}$ for ${\rm Br}_{\rm vis}\epsilon_{\rm rec}=1$. The quoted volumes are order-of-magnitude benchmarks; a detector-level study should replace them by topology- and energy-dependent effective volumes.
}
\label{tab:far_detector_comparison}
\end{table}

The {large number required} corresponds to a parent-particle production cross section of
\beq \sigma
\simeq
0.3~\mu{\rm b}
\left(\frac{N}{10^{11}}\right)
\left(\frac{320~{\rm fb}^{-1}}{\mathcal L_{\rm int}}\right).\eeq
This motivates a focus on hadronic interactions, {since hadron colliders produce enormous numbers of mesons.}

\begin{figure*}[p]
\centering
  \includegraphics[
    width=\linewidth,
    height=0.185\textheight,
    keepaspectratio
  ]{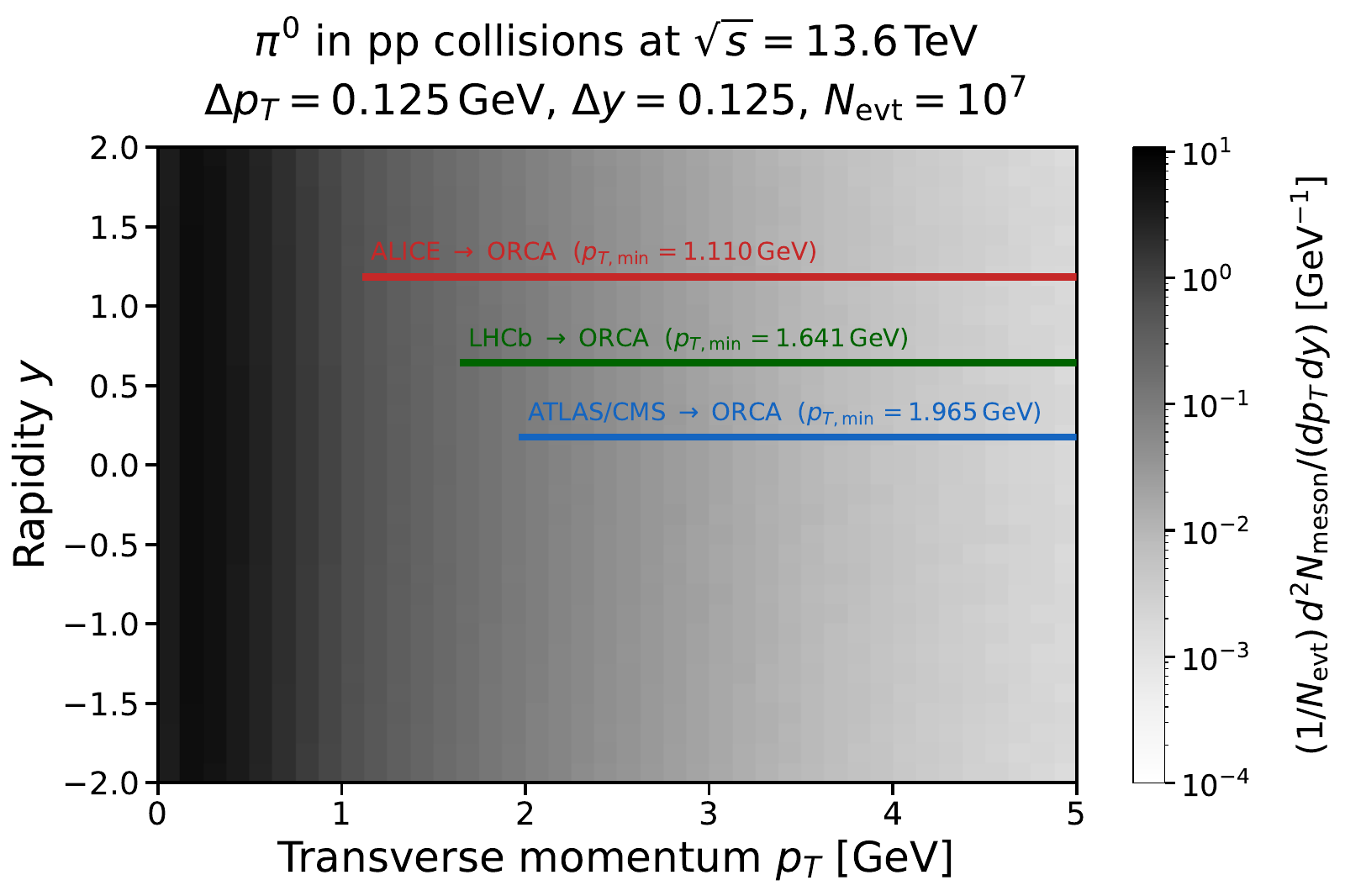}
  \includegraphics[
    width=\linewidth,
    height=0.185\textheight,
    keepaspectratio
  ]{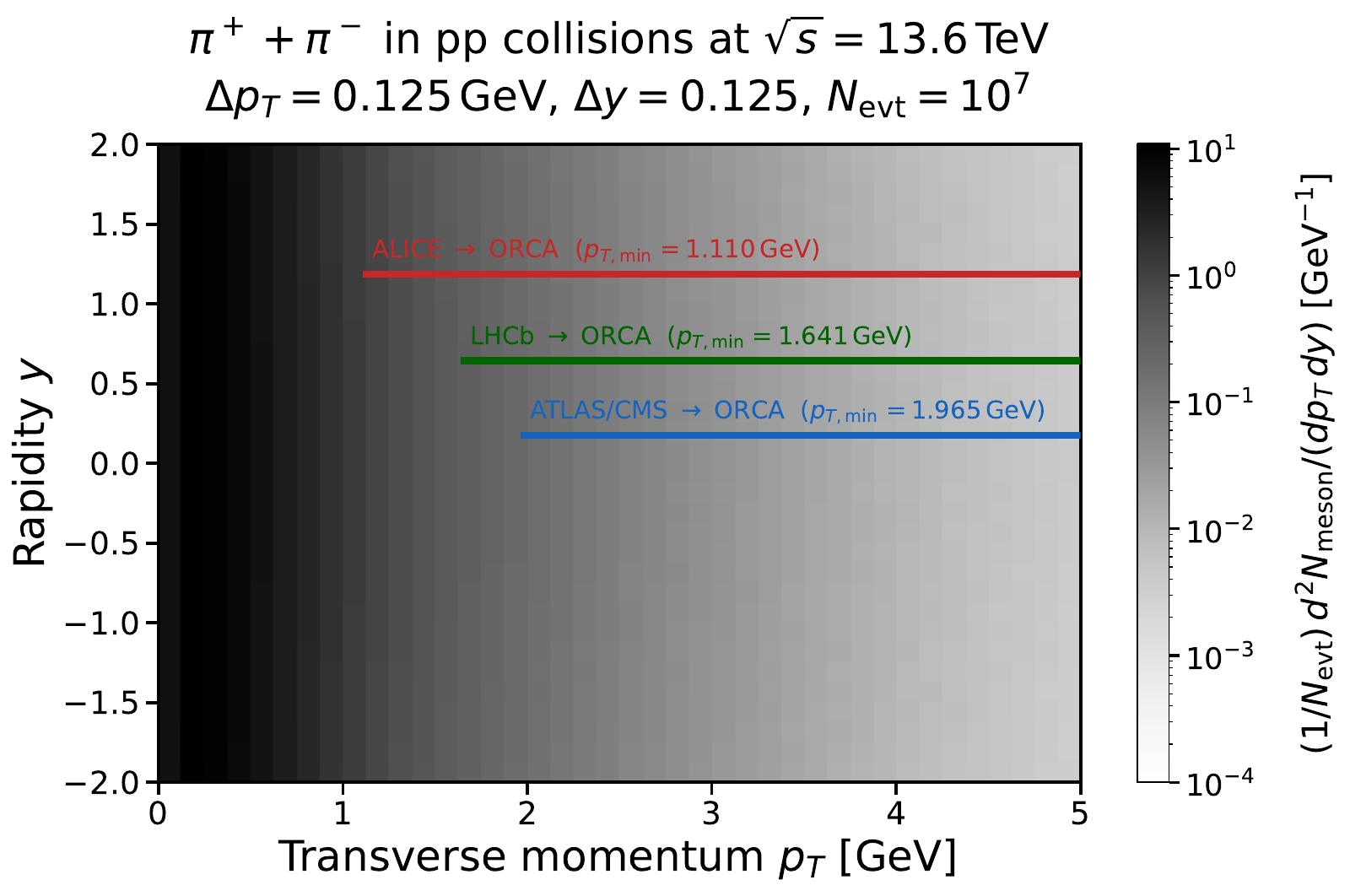}
  \includegraphics[
    width=\linewidth,
    height=0.185\textheight,
    keepaspectratio
  ]{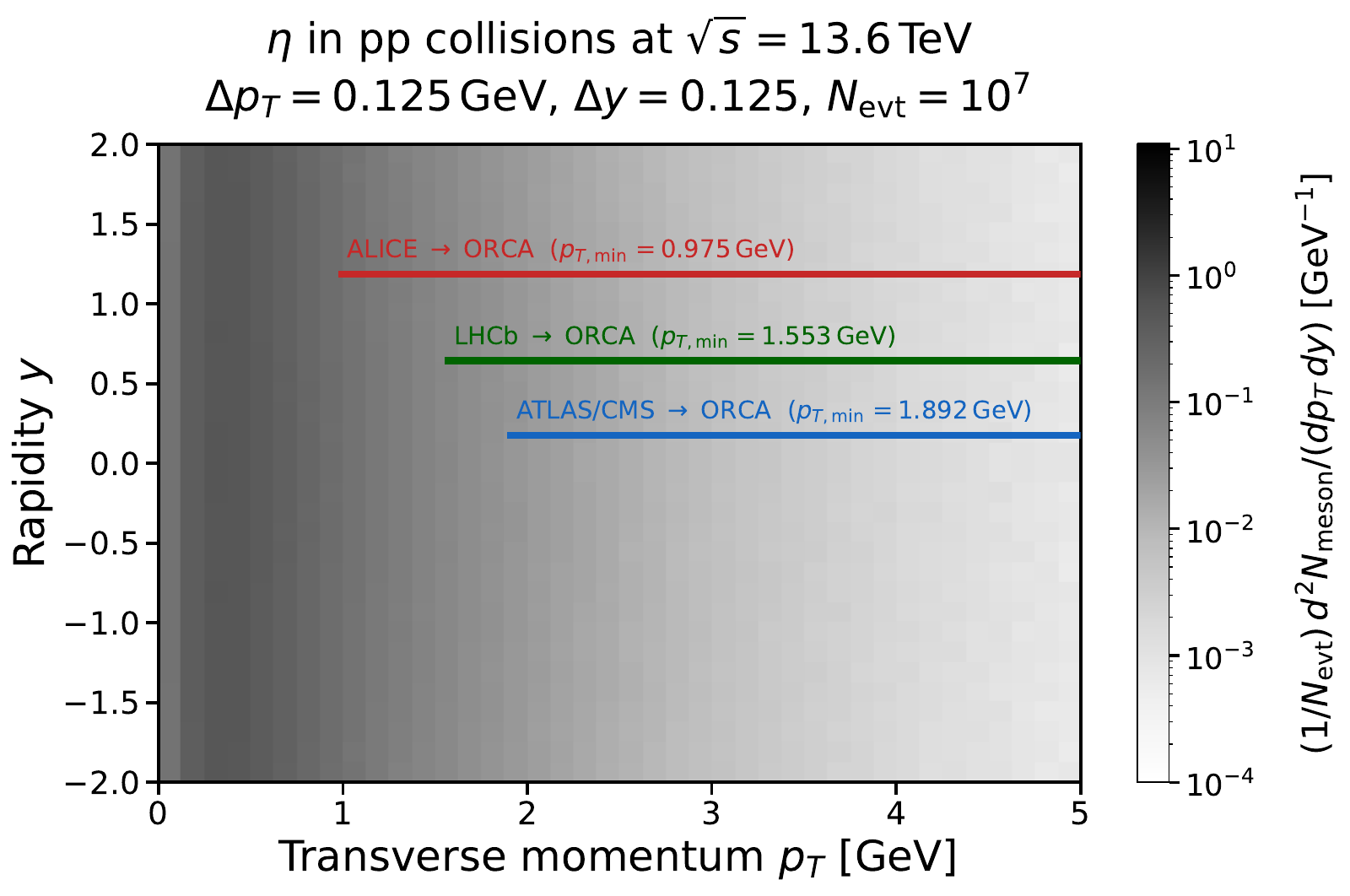}
  \includegraphics[
    width=\linewidth,
    height=0.185\textheight,
    keepaspectratio
  ]{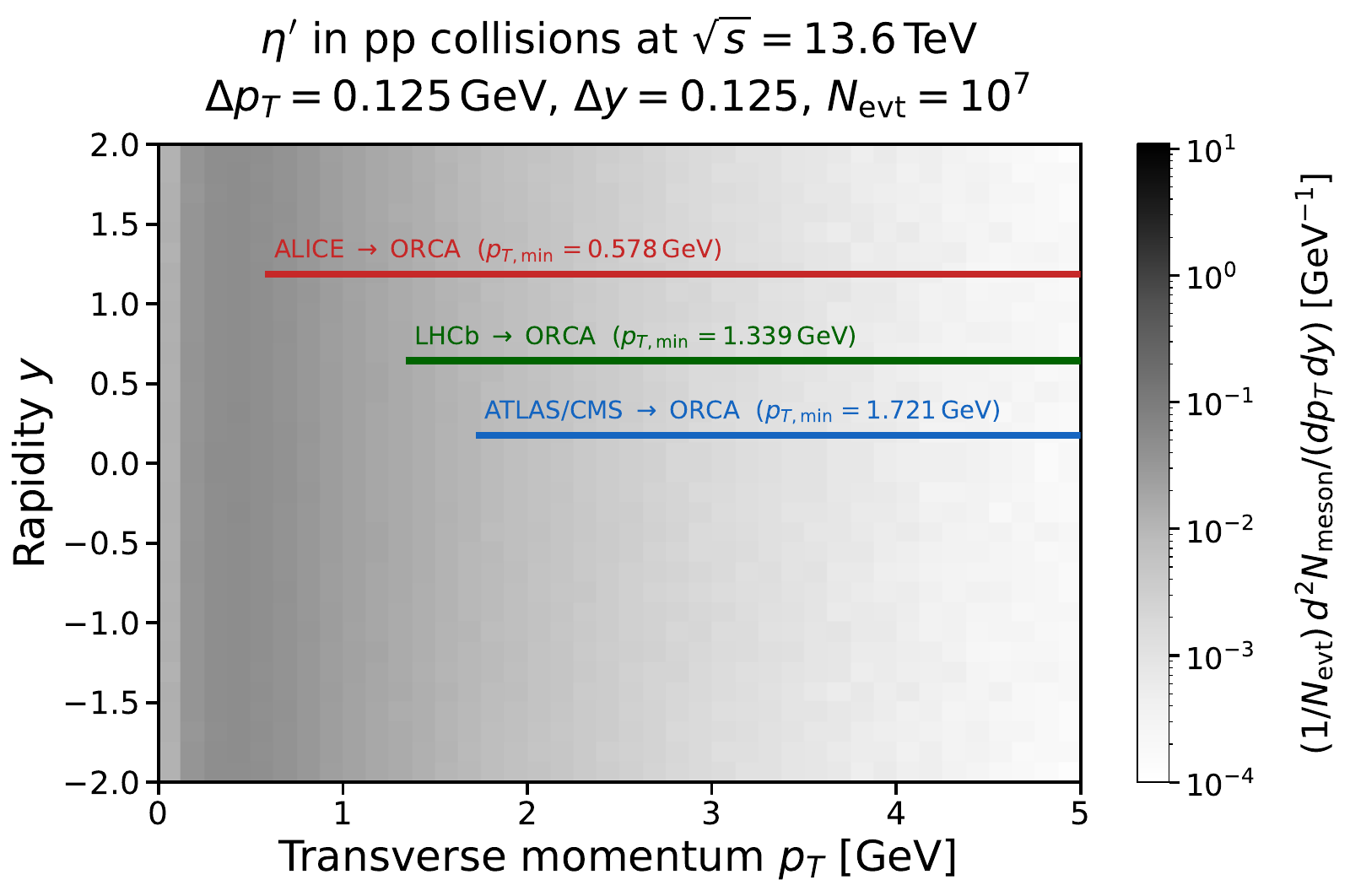}
  \includegraphics[
    width=\linewidth,
    height=0.185\textheight,
    keepaspectratio
  ]{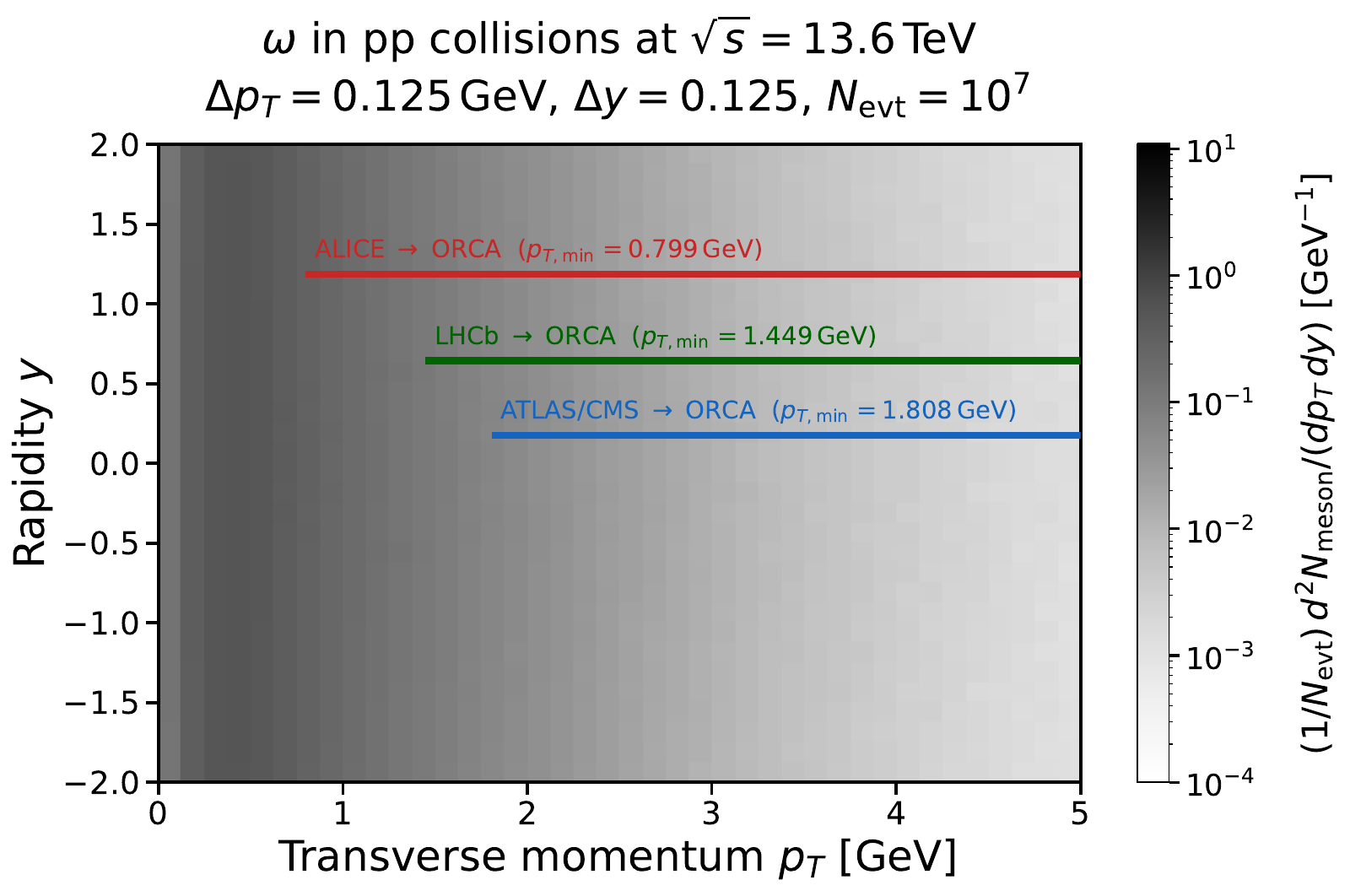}
  \includegraphics[
    width=\linewidth,
    height=0.185\textheight,
    keepaspectratio
  ]{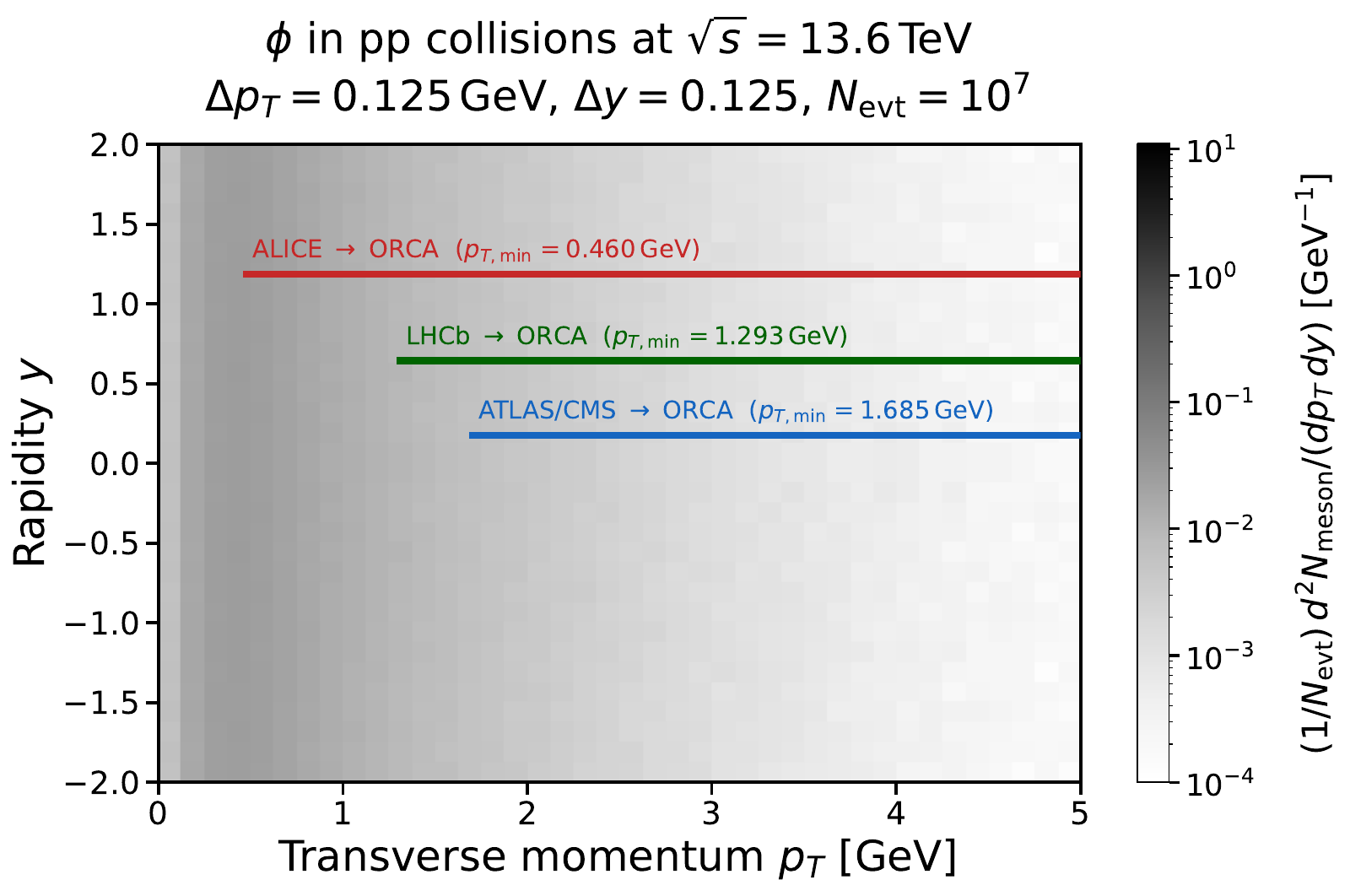}
  \includegraphics[
    width=\linewidth,
    height=0.185\textheight,
    keepaspectratio
  ]{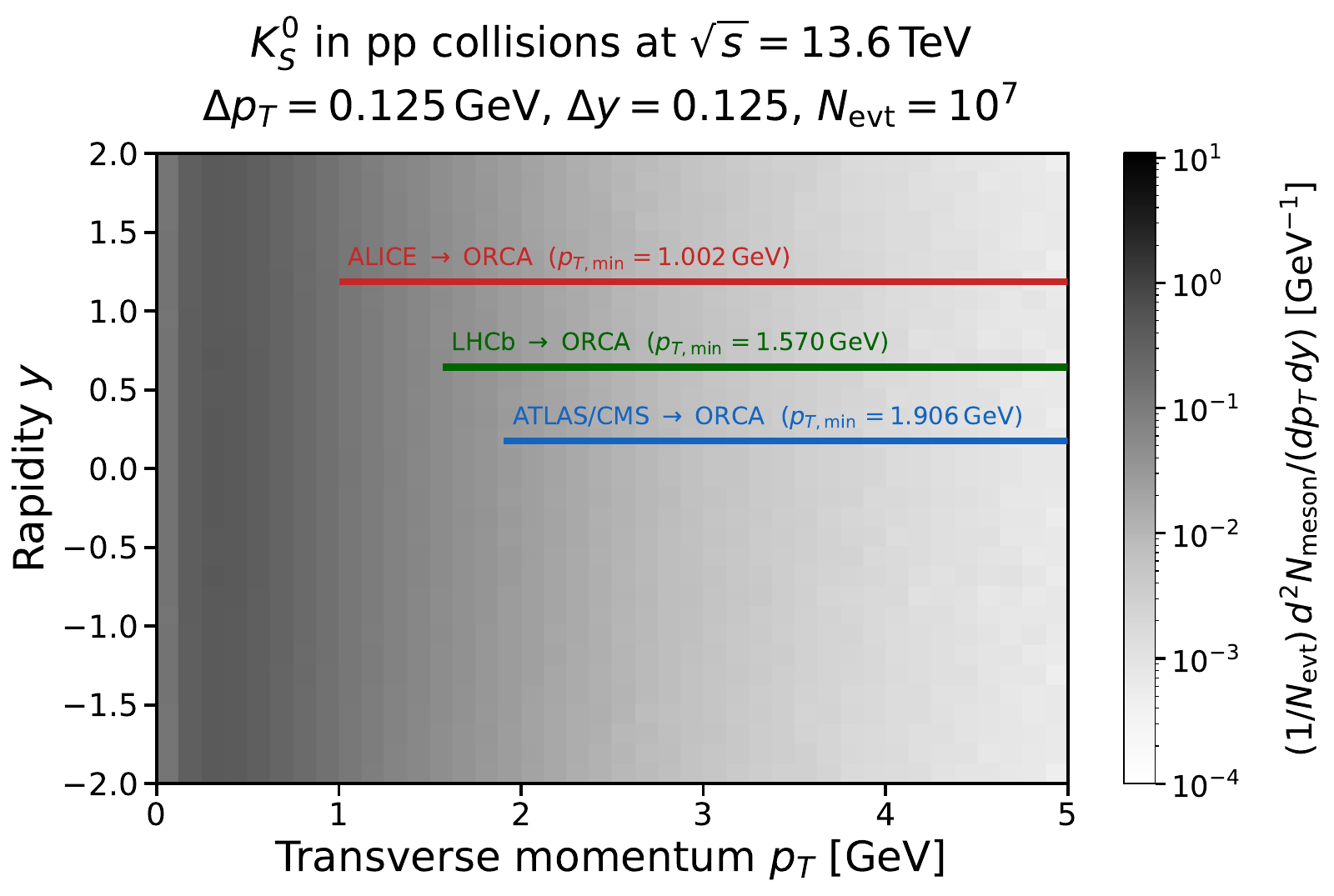}
  \includegraphics[
    width=\linewidth,
    height=0.185\textheight,
    keepaspectratio
  ]{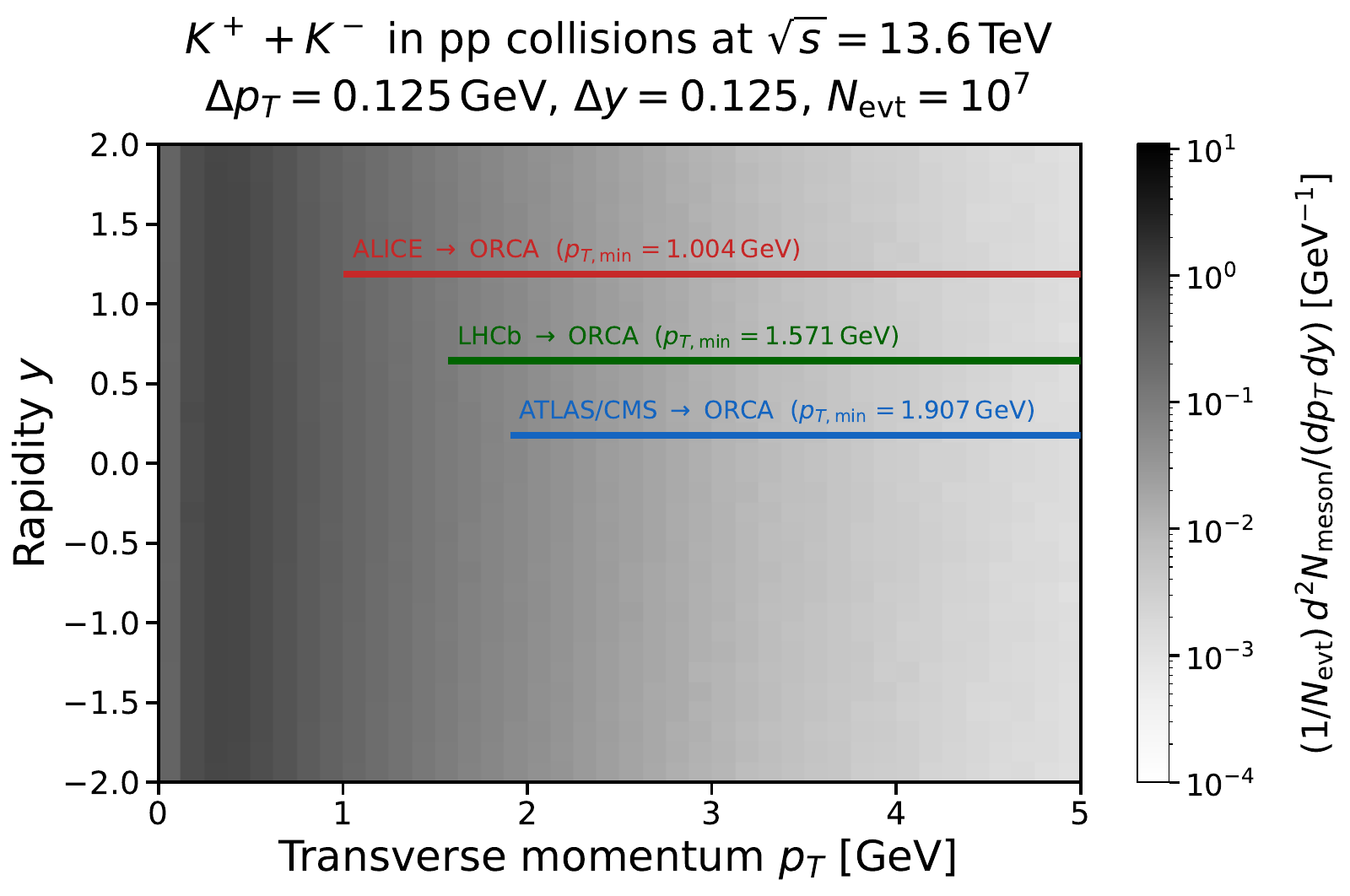}
\caption{
Generator-level $p_T-y$ density distributions
$\rho_{\rm meson}(p_T, y)$ for
$\pi^0$,~$\pi^++\pi^-$,~$\eta$,~$\eta^\prime$,~$\omega$,~$\phi$,
~$K_S^0$, and~$K^++K^-$
in inelastic $pp$ collisions at $\sqrt{s}=13.6~\mathrm{TeV}$, generated
with PYTHIA~8.
The red, green{,} and blue segments mark the approximate ALICE-to-ORCA, LHCb-to-ORCA{,} and ATLAS/CMS-to-ORCA {directions, respectively,} using $y\simeq|\eta_{\rm ORCA}|$.
{They also indicate} the species-dependent transverse-momentum threshold implied by
$E_{\rm meson}>2~\mathrm{GeV}$ for each meson.
}
\label{fig:meson-pt-y-density}
\end{figure*}

{To estimate meson production and hence $\partial_\Omega \F_X$ in the next section, we define, for each meson species, the normalized density}
\begin{equation}
\rho_{\rm meson}(p_T,y)
\equiv
\frac{1}{N_{\rm evt}}
\frac{d^2N_{\rm meson}}{dp_T\,dy},
\label{eq:meson-density}
\end{equation}
which gives the average meson multiplicity per inelastic $pp$ collision, per unit transverse momentum {and per unit rapidity}.
Fig.~\ref{fig:meson-pt-y-density} shows the generator-level transverse-momentum--rapidity density distributions of the meson species relevant for the production of long-lived particles through rare meson decays. 
The distributions were obtained from {PYTHIA~8.317 simulations of inelastic $pp$ collisions using SoftQCD:inelastic mode} at $\sqrt{s}=13.6~\mathrm{TeV}$ \cite{Bierlich:2022pfr}.\footnote{The generated data are checked to be consistent with the ALICE result at Run2~\cite{ALICE:2020jsh}.}
The exact forward--backward symmetry of the generator-level $pp$ initial state was used to average the $y$ and $-y$ bins. Thus, although the plots are normalized using an effective sample
size of $10^7$ events, only $5\times10^6$ statistically independent collisions were generated.  

All meson species exhibit a strong enhancement at small $p_T$, reflecting the dominance of soft hadron production in inclusive inelastic collisions. This motivates us to focus on ORCA, which is sensitive down to the GeV scale, rather than ARCA, which targets much higher energies.
{Also shown in Fig.~\ref{fig:meson-pt-y-density} are lines for the ALICE-to-ORCA, LHCb-to-ORCA, and ATLAS/CMS-to-ORCA directions (see also Table~\ref{tab:LHC_input}) and the typical ORCA energy sensitivity, for which we take $E_M>2\GEV$ for simplicity. When we estimate the flux in the next section, the lines dominate the integration.}

The overall normalization, however, depends substantially on the meson species. In particular, charged and neutral pions have the largest multiplicities, whereas the production rates of the heavier $\eta^\prime$,
$\omega$, and $\phi$ mesons are {lower}.

\section{Dark particles from rare meson decays: ORCA meets the LHC}
\lac{interpretation}

{According to the previous discussion, we focus on the possibility that $X$ is produced in meson decays at the LHC and search for its subsequent interaction in ORCA. To derive meaningful sensitivities, we study the kinematics and detector efficiency in detail.}

\paragraph{Kinematics} 

{The per-collision flux of $X$ via visible long-lived-particle decays can then be estimated as}
\begin{align}
\partial_\Omega {\Phi^{i}_X}
\simeq & \,\mathrm{BR}(M\to n_X X)\,n_X 
\int dp_T \, dy \, \frac{d\phi}{2\pi}
\frac{1}{N_{\rm evt}}\frac{d^2 N_M}{dp_T dy} \, 
\nonumber
\times 
K_i(p_T,y,\phi).
\label{eq:orca_optimized_yield}
\end{align}
Here $n_X$ is the multiplicity of the decay product $X$.
The total number of {$X$ particles} is $\sum_{i}\sigma_{\mathrm{inel}}\mathcal{L}_{\mathrm{int}}^{i}\partial_\Omega {\Phi^{i}_X} \times A_\perp/L_i^2$. {Here,} $\sigma_{\mathrm{inel}}\mathcal{L}_{\mathrm{int}}^{i}$ is the number
of inelastic collisions at interaction point $i$, {and $L_i$ is the distance from that interaction point to ORCA; these distances are almost the same. $K_i(p_T,y,\phi)$ is the lab-frame angular distribution for a meson $M$ to produce a decay product $X$ moving toward ORCA. For example, for a two-body meson decay to a relativistic $X$, it can be expressed as}
\begin{align}
    K_i(p_T,y,\phi)
    \equiv
    \frac{m^2_M}{4\pi \left(E_M - P_M \cos \psi_i \right)^2}~, ~~~~P_M\equiv \sqrt{E^2_M - m^2_M}
\end{align}
where $\psi_i$ is the lab-frame angle between the meson and an {$X$ particle directed toward ORCA; it satisfies}
\begin{align}
    \cos \psi_i
    =
    \frac{E_M \tanh y \, \sinh \eta_{\rm ORCA} + p_T \cos\phi }{P_M \, \cosh \eta_{\rm ORCA} }
\end{align}
with the psuedorapidity to ORCA $\eta_{\rm ORCA}$.

Then{,} for a given visible decay length $\lambda_X = \beta \gamma c \tau_X$, the expected number of
reconstructed LLP decays in KM3NeT-ORCA is
\begin{align}
N_{X}^{\mathrm{ORCA}}
= & \,
\mathrm{BR}_{M\to n_X X} 
\sum_i
\sigma_{\mathrm{inel}}\,
\mathcal{L}_{\mathrm{int}}^{i}
\,n_X 
\int dp_T \, dy \, \frac{d\phi}{2\pi}
\frac{1}{N_{\rm evt}}\frac{d^2 N_M}{dp_T dy}
\nonumber
\\ & \,
\times
K_i(p_T,y,\phi) \, 
\frac{V^{\mathrm{ORCA}}_{\rm eff} \exp \left(-L_i/\lambda_X \right)}
     {\,L_i^{\,2} \lambda_X}\,
\epsilon^\mathrm{ORCA}_{\mathrm{rec}}.
\label{eq:Nobs_BRproduct}
\end{align}
{Here,} the sum runs over the ATLAS, CMS, ALICE, and LHCb interaction points.
We {also} introduce the detector efficiency{, which is discussed below}.

\paragraph{Detector efficiency}
For $E_X \gtrsim 1\GEV$, {ORCA is sensitive to $X$ as long as it decays visibly; i.e., the $n$-body decay gives the same results.}
This is because the shower energy is
inferred from the aggregate number and spatial distribution of
photomultiplier hits.
{We therefore} impose the energy requirement on the total reconstructed visible
energy~\cite{Hofestadt:2015ctu},
\begin{equation}
E_{\rm vis}=\sum_{j\in {\rm visible}} E_j > E_{\rm cut}.
\end{equation}
{Assuming that all modes are visible, we set}
$E_{\rm vis}=E_X$ and consequently apply
$E_X>1~{\rm GeV}$ at the phenomenological level~\cite{KM3NeT:2024pte}. This approximation does
not require each individual daughter particle to have an energy above
$1~{\rm GeV}$. {We then} obtain the efficiency factor
\beq
\epsilon^\mathrm{ORCA}_{\mathrm{rec}}\equiv 
\Theta (E_{X,i} - E_{X,{\rm min}}).
\eeq 

For Run~3, {an additional factor should be taken into account}.
During Run~3, ORCA was still {being upgraded}, with the number of
deployed detection units (DUs) increasing throughout the run.\footnote{
See the KM3NeT--ORCA sea-campaign history:
\url{https://www.km3net.org/category/archive/orca-sea-campaigns/}.}
Since the completed ORCA detector has 115 DUs, we account for the
smaller Run~3 detector by a luminosity-weighted suppression factor
\begin{equation}
 f_{\rm ORCA}^{{\rm Run\,3}}
 =
 \frac{\sum_j {\cal L}_j N_{{\rm DU},j}}
      {115\sum_j {\cal L}_j}
 \simeq 0.22 ,
 \label{eq:ORCARun3factor}
\end{equation}
where ${\cal L}_j$ and $N_{{\rm DU},j}$ are the {accumulated LHC luminosity}
and the number of deployed ORCA DUs during period $j$, respectively.
We therefore multiply all Run~3 signal yields by
$f_{\rm ORCA}^{{\rm Run\,3}}=0.22$, while the completed ORCA configuration
is assumed for the HL-LHC projection.

\begin{figure*}[p]
\centering
  \includegraphics[
    width=\linewidth,
    height=0.75\textheight,
    keepaspectratio
  ]{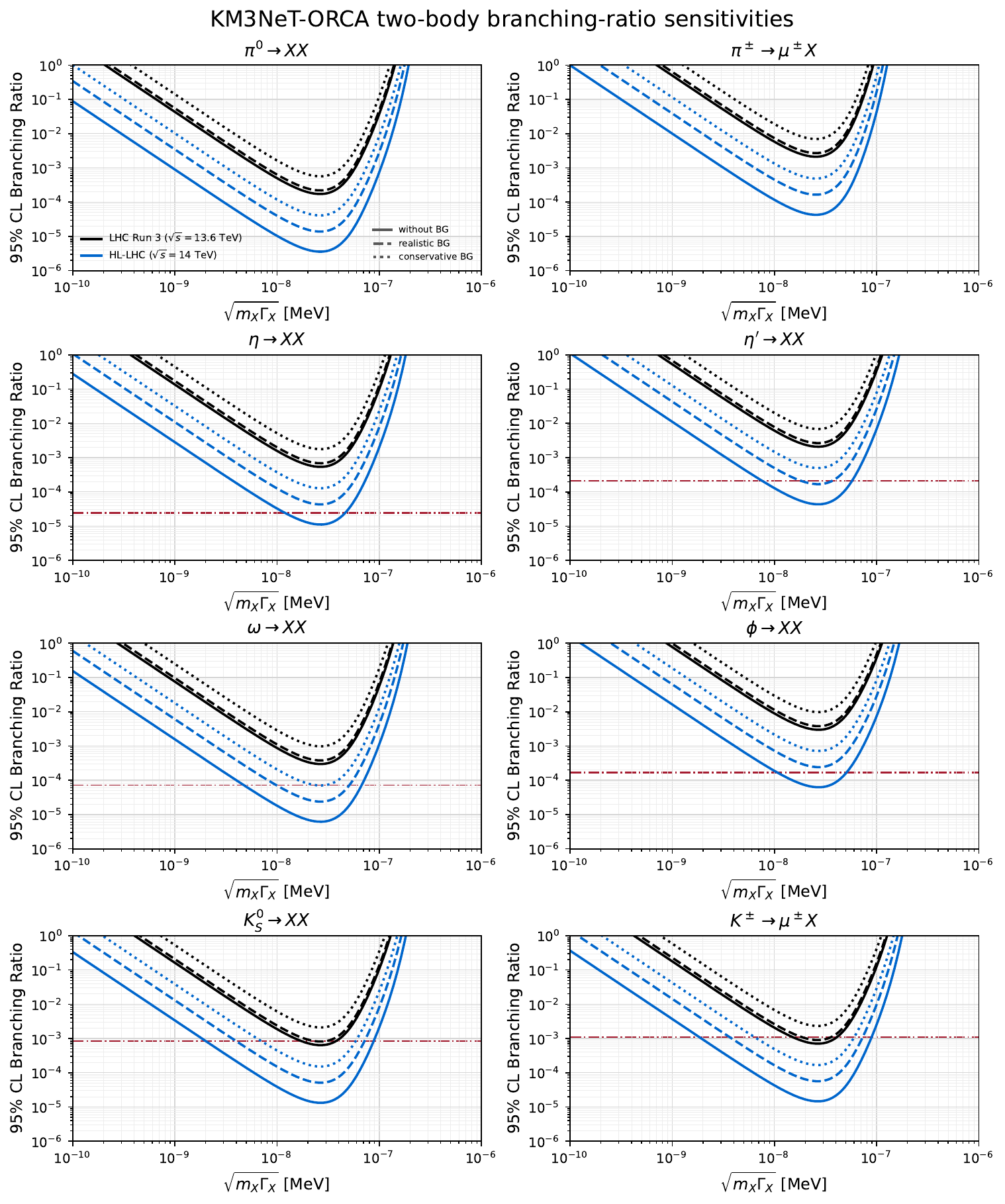}
\caption{
The projected $95\%$ C.L. sensitivity to
the meson {branching ratio} to $X$, $\mathrm{BR}_M$,
as a function of $\sqrt{m_X \Gamma_X}$ for LLPs produced in
rare decays of
$\pi^0$, $\pi^\pm$, $\eta$, $\eta^\prime$, $\omega$, $\phi$, $K_S^0$,
and $K^\pm$
through the decay modes $M\to XX$ for neutral mesons and $M\to\mu+X$ for charged mesons~\cite{ParticleDataGroup:2026mpi,BESIII:2025kjj}.
The solid lines are obtained by requiring
three reconstructed LLP decays in KM3NeT-ORCA,
assuming a background-free search.
The dashed (dotted) lines are given by the realistic (conservative) background scenarios.
The black (blue) curves correspond to the LHC Run~3 (HL-LHC) projections.
The horizontal dot-dashed line denotes the upper limit {on}
the branching ratio for each meson species.
}
\label{fig:orca_br_mgamma_sensitivity_with_and_without_BG}
\end{figure*}

\paragraph{Results}
The results are {shown} in Fig.~\ref{fig:orca_br_mgamma_sensitivity_with_and_without_BG} {as a function of} $\sqrt{m_X\Gamma_X}$, which is {a Lorentz-invariant} combination that can be probed ({cf.} $\lambda_X{(}P_X{)}=P_X/(m_X\Gamma_X)${,} where $P_X$ varies {from event to event}).
{The solid lines denote the $95\%$ C.L. Poisson upper limit {obtained} by neglecting the background, {i.e.,} $N^{\rm ORCA}_{X}=3$, while the other lines {use the background estimates} discussed in the next section. The Run~3 projections are displayed in black, whereas the HL-LHC
projections are shown in blue. 
For the HL-LHC scenario, we use the planned integrated luminosities of
$3~\mathrm{ab}^{-1}$ for both ATLAS and CMS,
$0.2\, \mathrm{fb}^{-1}$ for ALICE and $300\, \mathrm{fb}^{-1}$ for LHCb.}

All sensitivity curves exhibit a characteristic optimal point at
$\lambda_X\simeq3.8\times 10^2~\mathrm{km}$, comparable to the distances between the
LHC interaction points and ORCA.
For $\lambda_X\ll L_i$, most LLPs decay before reaching ORCA, leading to
the exponential suppression $\exp(-L_i/\lambda_X)$.
In the opposite limit, $\lambda_X\gg L_i$, the survival probability is
large, but the probability of decay inside the finite ORCA volume decreases
approximately as $1/\lambda_X$.
Thus the competition between these two effects produces the optimal sensitivity
at $\lambda_X\sim L_i$. 

Because the same reaction {appears as a missing-energy signature in conventional experiments~\cite{BESIII:2018bec,BESIII:2012nen,BESIII:2026qsu}, irrespective of} $m_X$,
we also show the existing limits {on} ${\rm BR}_{M}$ {as} horizontal dot-dashed lines. {In fact, for the $\pi$-decay case,} the limit is so strong that {the entire region} is excluded.
However, for other mesons{,} the HL-LHC can {provide better sensitivity}. LHC {Run~3 can provide a $95\%$ C.L. limit on kaon decays that is competitive with the missing-energy constraints}.

\begin{figure*}
\centering
  \includegraphics[
    width=0.5 \linewidth,
    keepaspectratio
  ]{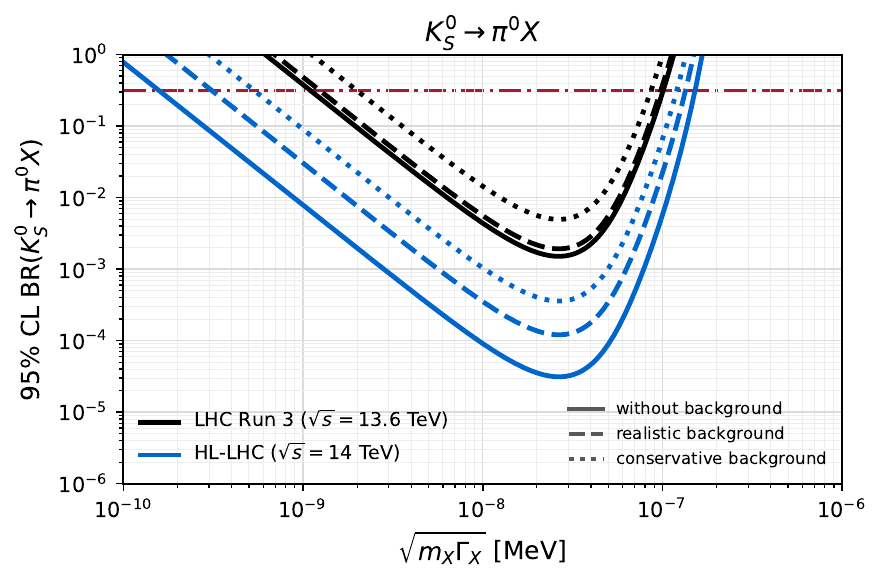}
\caption{
The projected $95\%$ C.L. sensitivity to the {branching ratio
$\mathrm{BR}(K_S^0\to\pi^0+X)$}
as a function of $\sqrt{m_X \Gamma_X}$ for LLPs produced in
rare decays of $K_S^0$.
The solid lines are obtained by requiring
three reconstructed LLP decays in KM3NeT-ORCA,
assuming a background-free search.
The dashed (dotted) lines are given by the realistic (conservative) background scenarios.
The black (blue) curves correspond to the LHC Run~3 (HL-LHC) projections. 
The horizontal dot-dashed line denotes the branching ratio {for the $\pi^0\pi^0$ decay mode,
${\rm BR}(K_S^0\to\pi^0\pi^0)=0.3069$, which is used as the upper limit on} the branching-ratio product. We take $f_{\rm time}=1$ because of the heavy $m_X$ (see the next section). 
}
\label{fig:orca_br_mgamma_sensitivity_K_S}
\end{figure*}

We also {note a potentially} large window {for $K_S^0\to\pi^0 X$} with $m_X\sim m_\pi$.
{This case is shown in Fig.~}\ref{fig:orca_br_mgamma_sensitivity_K_S}.
To our knowledge, no dedicated experimental limit on $K_S^0\to\pi^0X_{\rm inv}$ has been reported. {Thus, we use ${\rm BR}_{\rm M}<0.3$ as a conservative limit; this is the measured branching fraction  of $K_S^0\to\pi^0\pi^0$.}\footnote{ Nevertheless, KLOE found only $5.5\times10^5$ events with exactly two prompt photons among approximately $7\times10^8$ tagged $K_S^0$ decays~\cite{KLOE:2007rta}. For $m_X\simeq m_{\pi^0}$, a simple reinterpretation using the published photon efficiencies suggests $\mathrm{BR}(K_S^0\to\pi^0X_{\rm inv})\lesssim1.4\times10^{-3}$, or approximately $4\times10^{-4}$ after the QCAL veto. Since no signal-specific efficiency analysis or likelihood is available, we do not impose this estimate as a formal limit.}
If this conservative limit {applies, LHC Run~3} can already place {a} much more stringent limit.

\section{Background estimation at KM3NeT--ORCA}
\label{sec:orca_background}

Let us estimate the possible background.
The principal background sources for the search considered here are
\begin{equation}
  B_{\rm tot}
  =
  B_{\nu_{\rm atm}}
  +B_{\mu_{\rm atm}}
  +B_{\rm noise}
  +B_{\nu_{\rm LHC}},
  \label{eq:background_components}
\end{equation}
where the terms denote atmospheric-neutrino interactions,
atmospheric muons, optical and instrumental noise, and neutrinos
produced in LHC collisions, respectively.

For the full ORCA configuration, the trigger rates due to atmospheric
muons, optical noise, and atmospheric neutrinos are approximately
$50~{\rm Hz}$, $54~{\rm Hz}$, and $8~{\rm mHz}$, respectively~\cite{KM3NeT:2021ozk}. After the standard ORCA reconstruction
and background classifiers, approximately $6.6\times10^4$ upgoing
atmospheric-neutrino events per year remain~\cite{KM3NeT:2021ozk}, corresponding to a rate
of about $2~{\rm mHz}$.
For Run~3, we also use the reduced ORCA effective volume, $V^{\rm ORCA}_{\rm LHC\,Run\,3}\simeq0.22\,V^{\rm ORCA}_{\rm eff}$.

We consider representative stable-beam livetimes
\begin{equation}
  T_{\rm SB}^{\rm Run\,3}=2.0\times10^7~{\rm s},
  \qquad
  T_{\rm SB}^{\rm HL}=6.0\times10^7~{\rm s}.
  \label{eq:stable_beam_livetimes}
\end{equation}
The initial numbers of selected atmospheric-neutrino candidates are
therefore
\begin{equation}
  B_{\nu_{\rm atm}}^{\rm std}
  \simeq
  \begin{cases}
    8.0\times10^3, & {\rm Run~3},\\
    1.2\times10^5, & {\rm HL\mbox{-}LHC}.
  \end{cases}
  \label{eq:initial_neutrino_background}
\end{equation}
The following selections are then applied specifically to the
LHC--ORCA search.

\subsection{Atmospheric-neutrino background}

\subsubsection{Directional selection.}

Particles traveling from the LHC interaction points reach ORCA from
a direction approximately $1.35^\circ$ below the ORCA horizon. At
$E_\nu=10~{\rm GeV}$, the median ORCA angular resolution is approximately
$7^\circ$--$9^\circ$, with {poorer} resolution at lower energies
\cite{KM3NeT:2021ozk}. We use
\begin{equation}
  \psi_{\rm cut}=15^\circ
  \quad {\rm (realistic)},\qquad
  \psi_{\rm cut}=20^\circ
  \quad {\rm (conservative)}.
\end{equation}
For a background approximately uniform in azimuth over the upgoing
hemisphere, the corresponding directional factors are
\begin{equation}
  f_{\rm dir}
  =
  1-\cos\psi_{\rm cut}
  \simeq
  \begin{cases}
    0.034, & \psi_{\rm cut}=15^\circ,\\
    0.060, & \psi_{\rm cut}=20^\circ.
  \end{cases}
  \label{eq:f_direction}
\end{equation}
The precise angular acceptance for a directly produced
$X\to e^+e^-$ cascade has not been published and must ultimately be
obtained from a dedicated ORCA detector simulation.

\subsubsection{Energy selection.}

The $K_S^0\to XX$ signal near its optimal decay length is concentrated
at $E_X=\mathcal{O}(1)~{\rm GeV}$. In our signal calculation,
approximately $98.5\%$ of the accepted $K_S^0$ signal has
$1<E_X<3~{\rm GeV}$, and more than $99.9\%$ has
$1<E_X<5~{\rm GeV}$. We therefore use the experimentally more robust
window
\begin{equation}
  1~{\rm GeV}<E_{\rm rec}<5~{\rm GeV},
  \label{eq:energy_selection}
\end{equation}
which allows for energy smearing near the threshold.

The ORCA6 atmospheric-neutrino measurement provides a data-based
estimate of the corresponding background fraction. The unfolded
numbers of $\nu_\mu+\bar{\nu}_\mu$ events in the true-energy intervals
$1$--$5$, $5$--$20$, and $20$--$100~{\rm GeV}$ are
$194$, $540$, and $1055$, respectively
\cite{KM3NeT:2024ecf}. Thus, only
\begin{equation}
  f_E
  \simeq
  \frac{194}{194+540+1055}
  \simeq0.11
  \label{eq:f_energy}
\end{equation}
of the selected ORCA6 atmospheric-neutrino sample lies between
$1$ and $5~{\rm GeV}$. Contrary to the simple expectation that the
background peaks near $10~{\rm GeV}$, the selected ORCA6 event sample
is dominated by the $20$--$100~{\rm GeV}$ interval. We use
$f_E=0.11$ as a data-motivated proxy. The fraction should be updated
using the response of the completed ORCA detector.

\subsubsection{Shower-topology selection.}

The $X\to e^+e^-$ signal is expected to appear as a contained
shower-like event with no entering muon track. The ORCA6 simulation
after its standard event selection contains~\cite{KM3NeT:2024ecf}
\begin{equation}
  N_{\nu_e{\rm CC}}=552,\qquad
  N_{\nu_\mu{\rm CC}}=2958,\qquad
  N_{\nu_\tau{\rm CC}}=162,\qquad
  N_{\rm NC}=222 .
\end{equation}
The fraction not belonging to the dominant $\nu_\mu$ charged-current
sample is therefore
\begin{equation}
  \frac{552+162+222}{552+2958+162+222}
  \simeq0.24.
  \label{eq:non_numu_fraction}
\end{equation}
This gives an approximate lower limit on the atmospheric-neutrino
fraction that can survive a track veto. Allowing for imperfect
classification at low energy, we adopt
\begin{equation}
  f_{\rm sh}=0.30
  \quad {\rm (realistic)},\qquad
  f_{\rm sh}=0.50
  \quad {\rm (conservative)}.
  \label{eq:f_shower}
\end{equation}

\subsubsection{LHC timing selection.}

KM3NeT is designed to provide approximately $1~{\rm ns}$ synchronization
between its optical modules \cite{KM3NeT:2019shc}. The LHC has a nominal
$25~{\rm ns}$ bunch spacing and an average bunch-crossing rate of about
$30~{\rm MHz}$ \cite{Bruning:2004ej}. Assuming that the absolute CERN--ORCA
timing offset is calibrated with comparable precision, a
$5~{\rm ns}$ full timing window for the two dominant interaction points,
ATLAS and CMS, gives
\begin{equation}
  f_{\rm time}
  \simeq
  2(30~{\rm MHz})(5~{\rm ns})
  \simeq0.30.
  \label{eq:f_time_realistic}
\end{equation}
We use $f_{\rm time}=0.30$ in the realistic case and
$f_{\rm time}=1.00$ in the conservative case. This assumes that the LLP is
sufficiently relativistic or that its propagation delay has been
included in the predicted arrival time.

{We stress that this is a nontrivial requirement: a massive $X$ arrives
later than a massless neutrino by
$\Delta t\simeq(L/c)\,m_X^2/(2E_X^2)
\simeq6~\mu{\rm s}\,(m_X/100~{\rm MeV})^2(1~{\rm GeV}/E_X)^2$,
which is two orders of magnitude larger than the bunch spacing. 
The
delay can be predicted for each mass hypothesis from the reconstructed
$E_X$, but the attainable window is then set by the ORCA energy
resolution, $\delta(\Delta t)\simeq2\Delta t\,(\delta E_X/E_X)$, rather
than by the $1~{\rm ns}$ detector synchronization. A nanosecond window
is therefore realistic only for $m_X/E_X\lesssim3\times10^{-3}$, and
$f_{\rm time}$ should be relaxed accordingly for heavier $X$. Note that $f_{\rm time}$  is at most $1$.} \\

Combining these factors gives
\begin{equation}
  B_{\nu_{\rm atm}}
  =
  B_{\nu_{\rm atm}}^{\rm std}
  f_{\rm dir}f_E f_{\rm sh}f_{\rm time}.
  \label{eq:final_neutrino_background}
\end{equation}
Numerically,
\begin{align}
  B_{\nu_{\rm atm}}^{\rm Run\,3}
  &\simeq
  \begin{cases}
    2.7, & {\rm realistic},\\
    26.4, & {\rm conservative},
  \end{cases}
  \\
  B_{\nu_{\rm atm}}^{\rm HL}
  &\simeq
  \begin{cases}
    40.4, & {\rm realistic},\\
    396, & {\rm conservative}.
  \end{cases}
\end{align}
These estimates use only suppression factors that can be motivated
from published ORCA results. {Reducing the background to} strictly
$\mathcal{O}(1)$ would require additional electromagnetic-shower
discrimination that has not yet been demonstrated.

\subsection{Other background sources}

\subsubsection{Atmospheric-muon background.}

Atmospheric muons dominate the ORCA trigger rate, but are strongly
suppressed by the upgoing-direction requirement, containment, and
machine-learning event classification. In ORCA6, the atmospheric-muon
{Monte Carlo} sample is reduced from $2.6\times10^8$ events after the anti-noise
selection to only $23$ events after the final boosted decision tree selection. The
resulting atmospheric-muon contamination is below $0.6\%$
\cite{KM3NeT:2024ecf}. The completed-detector sensitivity study obtains
approximately $3\%$ contamination with only a $5\%$ loss of neutrino
efficiency \cite{KM3NeT:2021ozk}.

The present search additionally requires a contained low-energy
cascade, a direction compatible with the LHC, and an LHC timing
coincidence. No published residual rate exists for this exact
selection. We consequently assume
\begin{equation}
  B_{\mu_{\rm atm}}<0.1\;(1)
  \quad {\rm for~Run~3},
  \qquad
  B_{\mu_{\rm atm}}<0.3\;(3)
  \quad {\rm for~the~HL\mbox{-}LHC},
\end{equation}
where the values outside (inside) parentheses denote the realistic
(conservative) assumptions. This contribution must be checked using
near-horizon data sidebands.

\subsubsection{Optical-noise background.}

The dominant optical background is associated with $^{40}{\rm K}$
decays and bioluminescence. ORCA rejects these events using the number
of causally connected hits and reconstruction-quality variables. The
completed-detector study reports that the residual noise contribution
can be safely neglected {after application of its noise classifier}
\cite{KM3NeT:2021ozk}. Because no absolute residual count is
published for the present LLP selection, we assume
\begin{equation}
  B_{\rm noise}<0.1\;(1)
  \quad {\rm for~Run~3},
  \qquad
  B_{\rm noise}<0.3\;(3)
  \quad {\rm for~the~HL\mbox{-}LHC}.
\end{equation}
The coherent shower, directional, and LHC timing requirements should
provide additional rejection. These numerical bounds are assumptions,
not published ORCA predictions.

\subsubsection{LHC-produced neutrino background.}

Neutrinos produced in LHC collisions are potentially more problematic,
because they are correlated with both the LHC direction and collision
time. Directional and timing selections therefore do not reject them.
There is currently no dedicated calculation of the few-GeV neutrino
flux from the four LHC interaction points in the ORCA direction.

A simple geometrical estimate gives
\begin{equation}
 B_{\nu_{\rm LHC}}
 \sim
 N_{\rm inel}\,
 n_\nu\,
 \frac{A_{\rm ORCA}}{4\pi L^2}\,
 P_{\nu{\rm int}},
 \label{eq:lhc_neutrino_estimate}
\end{equation}
where $A_{\rm ORCA}\simeq\pi(115~{\rm m})^2$ and
$P_{\nu{\rm int}}\sim10^{-10}$ for a few-GeV neutrino traversing
$\mathcal{O}(200~{\rm m})$ of seawater. Taking the relevant but uncertain
neutrino multiplicity to be $n_\nu=\mathcal{O}(1$--$10)$ gives the provisional
ranges
\begin{equation}
  B_{\nu_{\rm LHC}}\sim0.1\text{--}1
  \quad {\rm for~Run~3},
  \qquad
  B_{\nu_{\rm LHC}}\sim1\text{--}10
  \quad {\rm for~the~HL\mbox{-}LHC}.
\end{equation}
These are order-of-magnitude estimates rather than literature-based
predictions.

\subsection{$K_S^0$ benchmark and sensitivity with nonzero background}
\label{sec:ks_background_sensitivity}

In the analytical sensitivity estimate we set
$\epsilon_{\rm sig}=1$. This choice should be interpreted as an
idealized upper-envelope sensitivity, in which
$V_{\rm eff}^{\rm ORCA}$ is understood as the effective volume after
triggering and reconstruction, and the energy, shower-topology, and
timing selections are chosen to retain the signal. In particular, the
$1$--$5~{\rm GeV}$ energy window retains more than $99.9\%$ of the
simulated $K_S^0$ signal near the optimal decay length, the
$X\to e^+e^-$ final state is intrinsically shower-like, and a timing
window centered on the predicted arrival time can have nearly unit
signal acceptance.

Nevertheless, unit efficiency cannot be established from
the available ORCA literature, because the angular and trigger
responses to a direct $X\to e^+e^-$ decay have not been simulated. If
the geometrical instrumented volume, rather than a post-selection
effective volume, is used for $V_{\rm eff}^{\rm ORCA}$, a separate
efficiency smaller than unity must be included. Thus,
$\epsilon_{\rm sig}=1$ is used as an idealized benchmark rather than
as a detector-performance prediction.

\begin{table}[t]
\centering
\begin{tabular}{lccc}
\hline\hline
Scenario
& $S$
& $B$
& $Z$
\\
\hline
Run~3, realistic
& $4.20$
& $3.0$
& $1.77$
\\
Run~3, conservative
& $4.20$
& $29.4$
& $0.74$
\\
HL-LHC, realistic
& $197$
& $42$
& $15.75$
\\
HL-LHC, conservative
& $197$
& $412$
& $8.48$
\\
\hline\hline
\end{tabular}
\caption{
Expected $K_S^0\to XX$ signal and total background at the optimal
decay parameter, using
${\rm BR}(K_S^0\to XX)=8.4\times10^{-4}$ and 
${\rm BR}(X\to{\rm vis})=1$.
The expected exclusion significance Z is calculated using the profile likelihood ratio approximation.
}
\label{tab:ks_signal_background_simple}
\end{table}

The ratio $S/B$ measures the purity of the selected sample, but it is
not a statistical significance and has no universal requirement for
a $95\%$ confidence-level exclusion. For a counting experiment with a
sufficiently large and precisely known background, the appropriate
simple approximation is
\beq 
 Z=\sqrt{2\left[S-B\ln \left(1+\frac{S}{B} \right)\right]}~.\eeq
A one-sided $95\%$ confidence-level exclusion approximately requires
$Z>1.64$. 

Under this approximation, the Run~3 $K_S^0$ benchmark is sensitive at
$95\%$ confidence level in the realistic case, whereas the conservative
case is slightly below the nominal threshold. The HL-LHC benchmark
remains sensitive in both cases.

\section{Conclusions and discussion}

{We have proposed a geometric search for very long-lived particles in which an existing or future large-volume detector serves as a remote decay volume for particles produced at a high-energy collider. The collider interaction point and the remote detector jointly determine the source--detector baseline, the signal direction, and the expected bunch-correlated arrival time. The decay probability is maximized when the laboratory-frame decay length is comparable to the baseline. Using meson production at the LHC and KM3NeT--ORCA as a benchmark, we have demonstrated the potential of this configuration to probe parameter regions complementary to those accessible with conventional collider detectors. The Run~3 and HL-LHC results presented here are sensitivity estimates for this new search strategy. {For the Run~3 case, we can already use the data to set a limit on very long displaced vertices that is competitive with the conventional missing-energy search. For future experiments, we expect much better sensitivity, especially with improved resolutions in energy, direction, timing, and topology.}}

{The same geometric method can be applied to future facilities, including FCC-ee, the ILC, and FCC-hh \cite{Abada:2019lih,Behnke:2013xla,ILC:2013jhg,FCC:2018vvp} as well as other future lepton and hadron colliders. At FCC-ee and the ILC, the controlled electron--positron initial state and well-defined bunch structure could enable complementary searches for LLPs produced in $Z$, Higgs, heavy-flavor, radiative, bremsstrahlung{,} or other exotic decays. Although the relevant production mechanisms differ from the LHC meson benchmark, the central principle remains unchanged: LLPs produced at a known interaction point are searched for through their decays in a distant detector located along a well-defined direction.}

Among these future colliders, FCC-hh is particularly important and should be regarded as a primary future target for this method. Its enormous hadronic event sample would provide an exceptionally intense source of mesons and other possible LLP parents, while its much higher collision energy would produce a substantial high-energy tail of strongly boosted LLPs capable of traversing very long baselines before decaying. The geometric search could therefore become an especially powerful complement to the conventional detector program at FCC-hh.

Moreover, an FCC-hh program would provide an opportunity to consider the interaction-point location, beam orientation, remote-detector position, and bunch-correlated timing together from the early design stage. Such a coordinated design could transform the presently accidental alignment between CERN and KM3NeT into a deliberately optimized long-baseline LLP facility involving a future large-volume neutrino telescope or another suitable remote detector. We therefore advocate including geometric LLP searches in the physics and infrastructure planning for FCC-hh, as well as in studies of FCC-ee, the ILC, and other future colliders.

\section*{Acknowledgments}
We would like to thank Kaoru Hagiwara {and} Alejandro Ibarra for useful discussions.
This work is supported by JSPS KAKENHI Grant Nos. 22K14029 (W.Y.),
23K22486 (W.Y.), and 26K00695 (W.Y.).  W.Y. is also supported by the Selective
Research Fund and the Incentive Research Fund of Tokyo Metropolitan University. Fig.~\ref{fig:image_geometric_detection} was created with the assistance of ChatGPT. The authors take full responsibility for its content.

\appendix 

\section{Geometry of the source--detector configuration}
\lac{appgeom}

Here we collect the elementary geometric relations that were used to
obtain the entries of Table~\ref{tab:geometry} and the last two columns
of Table~\ref{tab:LHC_input}. We approximate the Earth as a sphere of
radius $R_\oplus=6371\,{\rm km}$; its oblateness shifts the baselines
considered here by less than $0.1\%$ and is irrelevant at the present
level of accuracy.

A point at geographic latitude $\varphi$, longitude $\lambda_{\rm geo}$
and depth $d$ below the local surface has the position vector
\begin{equation}
  {\bf r}(\varphi,\lambda_{\rm geo},d)
  =
  (R_\oplus-d)
  \begin{pmatrix}
    \cos\varphi\cos\lambda_{\rm geo}\\
    \cos\varphi\sin\lambda_{\rm geo}\\
    \sin\varphi
  \end{pmatrix},
  \laq{rvec}
\end{equation}
and the central angle $\Delta$ subtended at the center of the Earth by
an interaction point (IP) and a detector $D$ follows from
\begin{equation}
  \cos\Delta
  =
  \sin\varphi_{\rm IP}\sin\varphi_D
  +\cos\varphi_{\rm IP}\cos\varphi_D
   \cos\!\left(\lambda_{\rm geo,IP}-\lambda_{{\rm geo},D}\right).
  \laq{centralangle}
\end{equation}
Writing $r_{\rm IP}=R_\oplus-d_{\rm IP}$ and $r_D=R_\oplus-d_D$, the
straight-line baseline through the Earth is
\begin{equation}
  L_D=\left|{\bf r}_D-{\bf r}_{\rm IP}\right|
  =\sqrt{r_{\rm IP}^{2}+r_D^{2}-2r_{\rm IP}r_D\cos\Delta}\,,
  \laq{baseline}
\end{equation}
with light-travel time $t_D=L_D/c$. It is $L_D$, the chord rather than
the surface arc, that enters the exponential suppression of
\Eq{probvolume}.

The elevation of the source as seen from the detector is
\begin{equation}
  \sin\alpha_D
  =
  \hat{\bf r}_D\cdot
  \frac{{\bf r}_{\rm IP}-{\bf r}_D}{\left|{\bf r}_{\rm IP}-{\bf r}_D\right|}
  =
  \frac{r_{\rm IP}\cos\Delta-r_D}{L_D}\,,
  \laq{elevation}
\end{equation}
where $\alpha_D<0$ means that the source lies below the local horizon of
the detector; the below-horizon angle quoted in
Table~\ref{tab:geometry} is $|\alpha_D|$. For $\Delta\ll1$ and
$d\ll R_\oplus$ this reduces to
\begin{equation}
  \alpha_D
  \simeq
  \frac{d_D-d_{\rm IP}}{L_D}-\frac{\Delta}{2}\,,
  \laq{elevationapprox}
\end{equation}
i.e., the source sits half a central angle below the horizon, partially
compensated by the depth of the detector. Because $\Delta$ is small for
a baseline of a few hundred kilometers, the CERN direction is only
marginally below the ORCA horizon, which is precisely why the rejection
of mis-reconstructed down-going atmospheric muons is less robust for
ORCA than for ARCA (Sec.~\ref{sec:orca_background}).

The deepest point of the chord fixes the amount of rock and water
traversed by $X$. The distance of closest approach of the segment
${\bf r}_{\rm IP}\to{\bf r}_D$ to the center of the Earth is
$p=r_{\rm IP}r_D\sin\Delta/L_D$, so that the maximum depth below the
reference surface is
\begin{equation}
  d_{\rm max}
  =
  R_\oplus-\frac{r_{\rm IP}r_D\sin\Delta}{L_D}
  \;\simeq\;
  \frac{L_D^{2}}{8R_\oplus}+\frac{d_{\rm IP}+d_D}{2}\,,
  \laq{chorddepth}
\end{equation}
the last form holding for $\Delta\ll1$. \Eq{chorddepth} applies as long
as the foot of the perpendicular lies between the two endpoints, which
is the case for all configurations considered here.

Inserting $L=3.82\times10^{2}\,{\rm km}$, $d_{\rm IP}\simeq0$ and
$d_D=2.45\,{\rm km}$ into \Eqs{elevation} and \eq{chorddepth} gives
$\alpha_{\rm ORCA}=-1.35^\circ$, $d_{\rm max}=4.2\,{\rm km}$ and
$t_D=1.27\,{\rm ms}$, while $L=1.385\times10^{3}\,{\rm km}$ with
$d_D=3.5\,{\rm km}$ gives $\alpha_{\rm ARCA}=-6.10^\circ$,
$d_{\rm max}=39.6\,{\rm km}$ and $t_D=4.62\,{\rm ms}$, reproducing
Table~\ref{tab:geometry}.

The same construction fixes the direction in which the meson spectra of
Fig.~\ref{fig:meson-pt-y-density} must be evaluated. Denoting by
$\theta_i$ the angle between the local LHC beam axis at interaction
point $i$ and the unit vector
$({\bf r}_{\rm ORCA}-{\bf r}_i)/L_i$, the ORCA direction corresponds to
the pseudorapidity
\begin{equation}
  \left|\eta_{\rm ORCA}^{(i)}\right|
  =-\ln\!\left[\tan(\theta_i/2)\right].
  \laq{etaorca}
\end{equation}
Since the four interaction points lie on a ring of only $27\,{\rm km}$
circumference, the baselines $L_i$ agree with one another to better
than $\sim 2\%$, and the approximation $L_i\simeq L$ used in
Sec.~\ref{chap:geometry} is well justified. The beam-axis orientations,
on the other hand, differ appreciably, so that
$|\eta_{\rm ORCA}^{(i)}|$ ranges from $0.17$ at ATLAS and CMS to
$1.19$ at ALICE. ATLAS and CMS thus view ORCA close to central
rapidity, where the meson densities in
Fig.~\ref{fig:meson-pt-y-density} are largest; together with their much
larger integrated luminosities this is why they dominate the yields of
Table~\ref{tab:Nobs_LHC}.

\section{Numerical data}
Some data used in the main parts are listed.

Table~\ref{tab:geometry} collects the site parameters of the two KM3NeT
building blocks together with the geometric quantities derived from them
using Appendix~\ref{chap:appgeom}. The quoted sizes are the instrumented
volumes and should be understood as order-of-magnitude benchmarks; a
detector-level analysis must replace them by energy- and
topology-dependent effective volumes.

Table~\ref{tab:LHC_input} lists, for each LHC interaction point, the
Run~3 integrated luminosity, the effective luminosity
$\mathcal{L}^{\rm eff}_{\rm Run3}=f_{\rm ORCA}^{\rm Run3}\mathcal{L}_{\rm Run3}$
that accounts for the partially deployed ORCA detector
(\Eq{ORCARun3factor}), the baseline to ORCA, and the beam-axis angle
$\theta_i$ with the corresponding $|\eta_{\rm ORCA}^{(i)}|$ of
\Eq{etaorca}. ATLAS and CMS together provide $96\%$ of the Run~3
integrated luminosity, so the sensitivity is driven by these two
interaction points.

Table~\ref{tab:Nobs_LHC} gives the expected number of reconstructed $X$
decays inside ORCA obtained from \Eq{Nobs_BRproduct}, evaluated at the
optimal decay parameter $\lambda_X\simeq L_i$ and for the branching
ratio quoted in the second column. For neutral mesons this branching
ratio is the upper limit on the invisible decay mode, assuming
$M\to XX$, while for charged mesons it is the uncertainty of the
dominant branching fraction, assuming $M^\pm\to\mu^\pm X$. The entries
scale linearly with ${\rm BR}_M$ and with ${\rm BR}(X\to{\rm vis})$, so
they can be rescaled to any other model assumption.

Table~\ref{tab:orca_background_summary} shows the cumulative effect of
the four selections of Sec.~\ref{sec:orca_background} on the
atmospheric-neutrino sample, followed by the three remaining
contributions, which are assumptions rather than published ORCA
predictions. Values outside (inside) parentheses correspond to the
realistic (conservative) scenario, and the last row is the total
background $B$ for the estimates of
Table~\ref{tab:ks_signal_background_simple}.

\begin{table}[t]
\centering
\begin{tabular}{lcc}
\hline\hline
 & KM3NeT/ORCA & KM3NeT/ARCA \\
\hline
Site {coordinates} & $42^\circ48'{\rm N},\,6^\circ02'{\rm E}$ & $36^\circ16'{\rm N},\,16^\circ06'{\rm E}$ \\
Depth & $2.45\,{\rm km}$ & $3.5\,{\rm km}$ \\
Completed size used here & $7\,{\rm Mton}\simeq7\times10^{-3}\,{\rm km}^3$ & $1.0\,{\rm km}^3$ \\
Baseline from CERN & $3.82\times10^2\,{\rm km}$ & $1.385\times10^3\,{\rm km}$ \\
Light-travel time & $1.27\,{\rm ms}$ & $4.62\,{\rm ms}$ \\
Below-horizon angle& $1.35^\circ$ & $6.10^\circ$ \\
Maximum chord depth & $4.2\,{\rm km}$ & $39.6\,{\rm km}$ \\
\hline\hline
\end{tabular}
\caption{Geometric inputs. The KM3NeT detector volumes and depths are taken from the KM3NeT design literature and status documents \cite{Adrian-Martinez:2016fdl}. In a detector-level study{,} the geometrical size should be replaced by a topology- and energy-dependent effective volume $V_{\rm eff}$. The completed ORCA volume is used for the HL-LHC projection. For Run~3, the time-dependent detector configuration during ORCA
construction is taken into account as described below. We approximate the Earth as a sphere with radius
$R_\oplus=6371\,{\rm km}.$ }
\label{tab:geometry}
\end{table}

\begin{table}[t]
\centering
\begin{tabular}{lccccc}
\hline\hline
Collision point
& {$\mathcal{L}_{\rm Run\,3}$} (
${\rm fb}^{-1}$)& {$\mathcal{L}_{\rm Run\,3}^{\rm eff}$} (
${\rm fb}^{-1}$)
& Distance to ORCA ($\mathrm{km}$)
& Angle to beam axis ($^\circ$)
& $\left| \eta_{\rm ORCA} \right|$
\\
\hline
ATLAS
& $332$
& $73.0$
& $381.6$
& $80.2$
& $0.172$
\\
CMS
& $326.6$
& $71.9$
& $389.9$
& $79.9$
& $0.178$
\\
ALICE
& $0.150$
& $0.033$
& $383.4$
& $34.0$
& $1.185$
\\
LHCb
& $27.89$
& $6.14$
& $382.3$
& $55.4$
& $0.644$
\\
\hline\hline
\end{tabular}
\caption{Run~3 integrated-luminosity and geometrical inputs for the four LHC interaction points used in the {KM3NeT--ORCA} projection.
The distance is the straight-line distance from each interaction point to ORCA, and the angle is {measured} between the local LHC beam axis and the direction toward ORCA.
The corresponding pseudorapidity is obtained from $|\eta_{\rm ORCA}|=-\ln[\tan(\theta/2)]$.
}
\label{tab:LHC_input}
\end{table}

\begin{table}[t]
\centering
\begin{tabular}{lccccccc}
\hline\hline
Meson
& ${\rm BR}_{{\rm meson} \to X}$
& $N^{\rm ORCA}_{X ,{\rm ATLAS}}$
& $N^{\rm ORCA}_{X ,{\rm CMS}}$
& $N^{\rm ORCA}_{X ,{\rm ALICE}}$
& $N^{\rm ORCA}_{X ,{\rm LHCb}}$
& $N^{\rm ORCA}_{X ,{\rm tot}}$
& $N^{\rm ORCA}_{X ,{\rm HL-LHC}}$
\\
\hline
$\pi^0$
& $4.4 \times 10^{-9}$
& $3.84 \times10^{-5}$
& $3.56 \times10^{-5}$
& $1.73 \times10^{-7}$
& $6.88 \times10^{-6}$
& $8.10 \times10^{-5}$
& $3.79 \times10^{-3}$
\\
$\pi^\pm$
& $4.0 \times 10^{-7}$
& $2.74 \times 10^{-4}$
& $2.54 \times 10^{-4}$
& $2.16 \times 10^{-6}$
& $5.97 \times 10^{-5}$
& $5.90 \times 10^{-4}$
& $2.79 \times 10^{-2}$
\\
$\eta$
& $2.4 \times 10^{-5}$
& $0.0675$
& $0.0625$
& $3.18 \times 10^{-4}$
& $0.0121$
& $0.142$
& $6.65$
\\
$\eta^\prime$
& $2.1 \times 10^{-4}$
& $0.152$
& $0.141$
& $6.64 \times 10^{-4}$
& $0.0292$
& $0.323$
& $15.0$
\\
$\omega$
& $7.3 \times 10^{-5}$
& $0.370$
& $0.343$
& $1.83 \times 10^{-3}$
& $0.0688$
& $0.783$
& $36.6$
\\
$\phi$
& $1.7 \times 10^{-4}$
& $0.0868$
& $0.0804$
& $3.28 \times 10^{-4}$
& $0.0163$
& $0.184$
& $8.53$
\\
$K^0_S$
& $8.4 \times 10^{-4}$
& $1.99$
& $1.85$
& $8.84 \times 10^{-3}$
& $0.351$
& $4.20$
& $197$
\\
$K^\pm$
& $1.1 \times 10^{-3}$
& $2.36$
& $2.18$
& $1.06 \times 10^{-2}$
& $0.418$
& $4.97$
& $233$
\\
\hline\hline
\end{tabular}
\caption{
Geometrically projected numbers of long-lived particles $X$ reaching KM3NeT--ORCA from rare meson {two-body} decays at the LHC.
The Run~3 results ({$N^{\rm ORCA}_{X,{\rm ATLAS}}$, $N^{\rm ORCA}_{X,{\rm CMS}}$, $N^{\rm ORCA}_{X,{\rm ALICE}}$, $N^{\rm ORCA}_{X,{\rm LHCb}}$, and $N^{\rm ORCA}_{X,{\rm tot}}$}) correspond to $pp$ collisions at $\sqrt{s}=13.6~\mathrm{TeV}$, while the last column gives the total HL-LHC projection at $\sqrt{s}=14~\mathrm{TeV}$ using the ATLAS/CMS integrated-luminosity benchmarks $\mathcal{L}_{\rm ATLAS/CMS} = 3~\mathrm{ab}^{-1}$ and $\mathcal{L}_{\rm ALICE+ LHCb}=(0.2+300)\rm fb^{-1}$.
Neutral-meson entries use the existing invisible-decay upper limits and assume two $X$ particles per decay, whereas charged-meson entries use the branching-ratio uncertainty and assume one $X$ particle per decay. The Run~3 estimate includes the reduced ORCA effective volume, $V^{\rm ORCA}_{\rm LHC\,Run\,3}\simeq0.22\,V^{\rm ORCA}_{\rm eff}$.
}
\label{tab:Nobs_LHC}
\end{table}

\begin{table*}[t]
\centering
\begin{tabular}{lccc}
\hline\hline
Background source/selection
& Relative factor
& LHC Run~3
& HL-LHC
\\
\hline
Atmospheric neutrino, {$B^{\rm std}_{\nu_{\rm atm}}$}
& ---
& $8.00\times10^{3}$
& $1.20\times10^{5}$
\\
Directional selection, $f_{\rm dir}$
& $0.034\;(0.060)$
& $2.72\times10^{2}\;(4.80\times10^{2})$
& $4.08\times10^{3}\;(7.20\times10^{3})$
\\
Energy selection, $f_E$
& $0.11\;(0.11)$
& $30.0\;(52.8)$
& $4.49\times10^{2}\;(7.92\times10^{2})$
\\
Topological selection, $f_{\rm sh}$
& $0.30\;(0.50)$
& $9.0\;(26.4)$
& $135\;(396)$
\\
LHC timing selection, $f_{\rm time}$
& $0.30\;(1.00)$
& $2.7\;(26.4)$
& $40.4\;(396)$
\\[1mm]
\hline
Atmospheric muon, {$B_{\mu_{\rm atm}}$}
& ---
& $\lesssim0.10\;(1.0)$
& $\lesssim0.30\;(3.0)$
\\[1mm]
\hline
Optical background, $B_{\rm noise}$
& ---
& $\lesssim0.10\;(1.0)$
& $\lesssim0.30\;(3.0)$
\\[1mm]
\hline
LHC-produced neutrino, {$B_{\nu_{\rm LHC}}$}
& ---
& $\sim0.1\;(1.0)$
& $\sim1.0\;(10.0)$
\\[1mm]
\hline
Sum
& ---
& $\lesssim 3.0\;(29.4)$
& $\lesssim42.0\;(412)$
\\
\hline\hline
\end{tabular}

\caption{
Estimated background event yields for the LHC--ORCA LLP search after
successive selections. Values outside (inside) parentheses correspond
to the realistic (conservative) assumptions. The atmospheric-neutrino
entries are cumulative, while the other contributions are approximate
benchmarks requiring dedicated detector simulations.
}
\label{tab:orca_background_summary}
\end{table*}
\clearpage
\bibliography{Geometric}
\end{document}